\batchmode
\documentstyle[12pt]{article}
\newcommand{\nc}{\newcommand}
\nc{\renc}{\renewcommand}

\nc{\half}{{\textstyle{1\over2}}}
\nc{\etal}{\mbox{\it et al. }}
\nc{\ie}{{\it i.e.}}
\nc{\eg}{{\it e.g.}}

\renc{\thefootnote}{\arabic{footnote}}
\nc{\capt}[1]{{\bf Figure.} {\small\sl #1}}

\nc{\eqs}[2]{\mbox{Eqs.~(\ref{#1},\,\ref{#2})}}
\nc{\eq}[1]{\mbox{Eq.~(\ref{#1})}}

\nc{\figs}[2]{\mbox{Figs.~(\ref{#1},\,\ref{#2})}}
\nc{\fig}[1]{\mbox{Fig~.(\ref{#1})}}

\nc{\tag}[1]{\label{#1} \marginpar{{\footnotesize #1}}}
\nc{\mtag}[1]{\label{#1} \mbox{\marginpar{{\footnotesize #1}}}}
\renc{\baselinestretch}{1.5}
\newlength{\overeqskip}
\newlength{\undereqskip}
\nc{\be}[1]{\begin{equation} \mbox{$\label{#1}$}}
\nc{\bea}[1]{\begin{eqnarray} \mbox{$\label{#1}$}}
\nc{\Section}[2]{\section{#2}\label{#1}}
\nc{\Bibitem}[1]{\bibitem{#1}}
\nc{\Label}[1]{\label{#1}}

\nc{\eea}{\vspace{\undereqskip}\end{eqnarray}}
\nc{\ee}{\vspace{\undereqskip}\end{equation}}
\nc{\bdm}{\begin{displaymath}}
\nc{\edm}{\end{displaymath}}
\nc{\dpsty}{\displaystyle}
\nc{\bc}{\begin{center}}
\nc{\ec}{\end{center}}
\nc{\ba}{\begin{array}}
\nc{\ea}{\end{array}}
\nc{\bab}{\begin{abstract}}
\nc{\eab}{\end{abstract}}
\nc{\btab}{\begin{tabular}}
\nc{\etab}{\end{tabular}}
\nc{\bit}{\begin{itemize}}
\nc{\eit}{\end{itemize}}
\nc{\ben}{\begin{enumerate}}
\nc{\een}{\end{enumerate}}
\nc{\bfig}{\begin{figure}}
\nc{\efig}{\end{figure}}
\nc{\arreq}{&\!=\!&}
\nc{\arrmi}{&\!-\!&}
\nc{\arrpl}{&\!+\!&}
\nc{\arrap}{&\!\!\!\approx\!\!\!&}
\nc{\non}{\nonumber\\*}
\nc{\align}{\!\!\!\!\!\!\!\!&&}

\def\lsim{\; \raise0.3ex\hbox{$<$\kern-0.75em
      \raise-1.1ex\hbox{$\sim$}}\; }
\def\gsim{\; \raise0.3ex\hbox{$>$\kern-0.75em
      \raise-1.1ex\hbox{$\sim$}}\; }
\nc{\DOT}{\hspace{-0.08in}{\bf .}\hspace{0.1in}}
\nc{\Laada}{\hbox {$\sqcap$ \kern -1em $\sqcup$}}
\nc\loota{{\scriptstyle\sqcap\kern-0.55em\hbox{$\scriptstyle\sqcup$}}}
\nc\Loota{{\sqcap\kern-0.65em\hbox{$\sqcup$}}}
\nc\laada{\Loota}
\nc{\qed}{\hskip 3em \hbox{\BOX} \vskip 2ex}

\nc{\real}{{\rm I \! R}}
\nc{\Z}{{\sf Z \!\!\! Z}}
\nc{\complex}{{\rm C\!\!\! {\sf I}\,\,}}
\def\bigid{\leavevmode\hbox{\small1\kern-3.8pt\normalsize1}}
\def\id{\leavevmode\hbox{\small1\kern-3.3pt\normalsize1}}
\nc{\slask}{\!\!\!/}
\nc{\bis}{{\prime\prime}}
\nc{\pa}{\partial}
\nc{\na}{\nabla}
\nc{\ra}{\rangle}
\nc{\la}{\langle}
\nc{\goto}{\rightarrow}
\nc{\swap}{\leftrightarrow}

\nc{\EE}[1]{ \mbox{$\cdot10^{#1}$} }
\nc{\abs}[1]{\left|#1\right|}
\nc{\at}[2]{\left.#1\right|_{#2}}
\nc{\norm}[1]{\|#1\|}
\nc{\abscut}[2]{\Abs{#1}_{\scriptscriptstyle#2}}
\nc{\vek}[1]{{\rm\bf #1}}
\nc{\integral}[2]{\int\limits_{#1}^{#2}}
\nc{\inv}[1]{\frac{1}{#1}}
\nc{\dd}[2]{{{\partial #1}\over{\partial #2}}}
\nc{\ddd}[2]{{{{\partial}^2 #1}\over{\partial {#2}^2}}}
\nc{\dddd}[3]{{{{\partial}^2 #1}\over
	{\partial #2 \partial #3}}}
\nc{\dder}[2]{{{d #1}\over{d #2}}}
\nc{\ddder}[2]{{{d^2 #1}\over{d {#2}^2}}}
\nc{\dddder}[3]{{d^2 #1}\over
	{d #2 d #3}}
\nc{\dx}[1]{d\,^{#1}x}
\nc{\dy}[1]{d\,^{#1}y}
\nc{\dz}[1]{d\,^{#1}z}
\nc{\dl}[1]{\frac{d\,^{#1}l}{(2\pi)^{#1}}}
\nc{\dk}[1]{\frac{d\,^{#1}k}{(2\pi)^{#1}}}
\nc{\dq}[1]{\frac{d\,^{#1}q}{(2\pi)^{#1}}}

\nc{\cc}{\mbox{$c.c.$ }}
\nc{\hc}{\mbox{$h.c.$ }}
\nc{\cf}{cf.\ }
\nc{\erfc}{{\rm erfc}}
\nc{\Tr}{{\rm Tr\,}}
\nc{\tr}{{\rm tr\,}}
\nc{\pol}{{\rm pol}}
\nc{\sign}{{\rm sign}}
\nc{\bfT}{{\bf T }}
\def\eV{{\rm\ eV}}
\def\GeV{{\rm\ GeV}}

\nc{\cA}{{\cal A}}
\nc{\cB}{{\cal B}}
\nc{\cD}{{\cal D}}
\nc{\cE}{{\cal E}}
\nc{\cG}{{\cal G}}
\nc{\cH}{{\cal H}}
\nc{\cL}{{\cal L}}
\nc{\cO}{{\cal O}}
\nc{\cT}{{\cal T}}
\nc{\cN}{{\cal N}}
\nc{\rvac}[1]{|{\cal O}#1\rangle}
\nc{\lvac}[1]{\langle{\cal O}#1|}
\nc{\rvacb}[1]{|{\cal O}_\beta #1\rangle}
\nc{\lvacb}[1]{\langle{\cal O}_\beta #1 |}
\nc{\bb}{\bar{\beta}}
\nc{\bt}{\tilde{\beta}}
\nc{\ctH}{\tilde{\cal H}}
\nc{\chH}{\hat{\cal H}}
\nc{\1}{\aa}
\nc{\2}{\"{a}}
\nc{\3}{\"{o}}
\nc{\4}{\AA}
\nc{\5}{\"{A}}
\nc{\6}{\"{O}}
\nc{\al}{\alpha}
\nc{\g}{\gamma}
\nc{\Del}{\Delta}
\nc{\e}{\epsilon}
\nc{\eps}{\epsilon}
\nc{\lam}{\lambda}
\nc{\om}{\omega}
\nc{\Om}{\Omega}
\nc{\ve}{\varepsilon}
\nc{\mn}{{\mu\nu}}
\nc{\k}{\kappa}
\nc{\vp}{\varphi}

\nc{\advp}[3]{{\it  Adv.\ in\ Phys.\ }{{\bf #1} {(#2)} {#3}}}
\nc{\annp}[3]{{\it  Ann.\ Phys.\ (N.Y.)\ }{{\bf #1} {(#2)} {#3}}}
\nc{\apl}[3]{{\it  Appl. Phys. Lett. }{{\bf #1} {(#2)} {#3}}}
\nc{\apj}[3]{{\it  Ap.\ J.\ }{{\bf #1} {(#2)} {#3}}}
\nc{\apjl}[3]{{\it  Ap.\ J.\ Lett.\ }{{\bf #1} {(#2)} {#3}}}
\nc{\app}[3]{{\it Astropart.\ Phys.\ }{{\bf #1} {(#2)} {#3}}}
\nc{\cmp}[3]{{\it  Comm.\ Math.\ Phys.\ }{{ \bf #1} {(#2)} {#3}}}
\nc{\cqg}[3]{{\it  Class.\ Quant.\ Grav.\ }{{\bf #1} {(#2)} {#3}}}
\nc{\epl}[3]{{\it  Europhys.\ Lett.\ }{{\bf #1} {(#2)} {#3}}}
\nc{\ijmp}[3]{{\it Int.\ J.\ Mod.\ Phys.\ }{{\bf #1} {(#2)} {#3}}}
\nc{\ijtp}[3]{{\it Int.\ J.\ Theor.\ Phys.\ }{{\bf #1} {(#2)} {#3}}}
\nc{\jmp}[3]{{\it  J.\ Math.\ Phys.\ }{{ \bf #1} {(#2)} {#3}}}
\nc{\jpa}[3]{{\it  J.\ Phys.\ A\ }{{\bf #1} {(#2)} {#3}}}
\nc{\jpc}[3]{{\it  J.\ Phys.\ C\ }{{\bf #1} {(#2)} {#3}}}
\nc{\jap}[3]{{\it J.\ Appl.\ Phys.\ }{{\bf #1} {(#2)} {#3}}}
\nc{\jpsj}[3]{{\it J.\ Phys.\ Soc.\ Japan\ }{{\bf #1} {(#2)} {#3}}}
\nc{\lmp}[3]{{\it Lett.\ Math.\ Phys.\ }{{\bf #1} {(#2)} {#3}}}
\nc{\mpl}[3]{{\it  Mod.\ Phys.\ Lett.\ }{{\bf #1} {(#2)} {#3}}}
\nc{\ncim}[3]{{\it  Nuov.\ Cim.\ }{{\bf #1} {(#2)} {#3}}}
\nc{\np}[3]{{\it  Nucl.\ Phys.\ }{{\bf #1} {(#2)} {#3}}}
\nc{\npps}[3]{{\it  Nucl.\ Phys.\ Proc.\ Suppl.\ }{{\bf #1} {(#2)} {#3}}}
\nc{\pr}[3]{{\it Phys.\ Rev.\ }{{\bf #1} {(#2)} {#3}}}
\nc{\pra}[3]{{\it  Phys.\ Rev.\ A\ }{{\bf #1} {(#2)} {#3}}}
\nc{\prb}[3]{{\it  Phys.\ Rev.\ B\ }{{{\bf #1} {(#2)} {#3}}}}
\nc{\prc}[3]{{\it  Phys.\ Rev.\ C\ }{{\bf #1} {(#2)} {#3}}}
\nc{\prd}[3]{{\it  Phys.\ Rev.\ D\ }{{\bf #1} {(#2)} {#3}}}
\nc{\prl}[3]{{\it Phys.\ Rev.\ Lett.\ }{{\bf #1} {(#2)} {#3}}}
\nc{\pl}[3]{{\it  Phys.\ Lett.\ }{{\bf #1} {(#2)} {#3}}}
\nc{\prep}[3]{{\it Phys.\ Rep.\ }{{\bf #1} {(#2)} {#3}}}
\nc{\prsl}[3]{{\it Proc.\ R.\ Soc.\ London\ }{{\bf #1} {(#2)} {#3}}}
\nc{\ptp}[3]{{\it  Prog.\ Theor.\ Phys.\ }{{\bf #1} {(#2)} {#3}}}
\nc{\ptps}[3]{{\it  Prog\ Theor.\ Phys.\ suppl.\ }{{\bf #1} {(#2)} {#3}}}
\nc{\physa}[3]{{\it  Physica\ A\ }{{\bf #1} {(#2)} {#3}}}
\nc{\physb}[3]{{\it  Physica\ B\ }{{\bf #1} {(#2)} {#3}}}
\nc{\phys}[3]{{\it Physica\ }{{\bf #1} {(#2)} {#3}}}
\nc{\rmp}[3]{{\it  Rev.\ Mod.\ Phys.\ }{{\bf #1} {(#2)} {#3}}}
\nc{\rpp}[3]{{\it Rep.\ Prog.\ Phys.\ }{{\bf #1} {(#2)} {#3}}}
\nc{\sjnp}[3]{{\it Sov.\ J.\ Nucl.\ Phys.\ }{{\bf #1} {(#2)} {#3}}}
\nc{\spjetp}[3]{{\it Sov.\ Phys.\ JETP\ }{{\bf #1} {(#2)} {#3}}}
\nc{\yf}[3]{{\it Yad.\ Fiz.\ }{{\bf #1} {(#2)} {#3}}}
\nc{\zetp}[3]{{\it Zh.\ Eksp.\ Teor.\ Fiz.\  }{{\bf #1}  {(#2)} {#3}}}
\nc{\zp}[3]{{\it Z.\ Phys.\ }{{\bf #1} {(#2)} {#3}}}
\nc{\ibid}[3]{{\sl ibid.\ }{{\bf #1} {#2} {#3}}}
\nc{\rf}[1]{(\ref{#1})}
\nc{\nn}{\nonumber \\*}
\nc{\bfB}{\bf{B}}
\nc{\bfv}{\bf{v}}
\nc{\bfx}{\bf{x}}
\nc{\bfy}{\bf{y}}
\nc{\vx}{\vec{x}}
\nc{\vy}{\vec{y}}
\nc{\oB}{\overline{B}}
\nc{\oI}{\overline{I}}
\nc{\oR}{\overline{R}}
\nc{\rar}{\rightarrow}
\nc{\ti}{\times}
\nc{\slsh}{\hskip-5pt/}
\nc{\sm}{Standard~Model~}
\nc{\MP}{M_{\rm Pl}}
\nc{\tp}{t_{\rm Pl}}
\nc{\ave}{\bar{E}}

\nc{\eff}{{\rm eff}}
\nc{\kk}{\vek{k}}
\nc{\pp}{{\rm p}}
\nc{\ga}{g_{a\gamma}}
\nc{\vv}{\\}
\nc{\eee}{{\bf E}}
\nc{\bbb}{{\bf B}}
\nc{\qcd}{T_{\rm QCD}}
\nc{\G}{\rm \ G}
\def\vec#1{{\bf #1}}

\def\lae{\;^{<}_{\sim} \;}  

\def\ell{e^{c}LL}

\begin{document}
{\title{\vskip-2truecm{\hfill {{\small \\
	\hfill \\
	}}\vskip 1truecm}
{\LARGE  Dynamical Cosmological Constant From A Very Recent Phase Transition}}
{\author{
{\sc  John McDonald$^{1}$}\\
{\sl\small Department of Physics and Astronomy,
University of Glasgow, Glasgow G12 8QQ, SCOTLAND}
}
\maketitle
\begin{abstract}
\noindent
           
          Observation indicates that the expansion of the Universe is accelerating and favours a
 dynamical cosmological constant, $\Lambda(t)$. We consider the possibility that this is due
 to a scalar field which has undergone a very recent phase transition. 
We study a simple class of model, corresponding to a $\phi^{4}$ potential with a time-dependent mass squared term.
 For the models considered the phase transition occurs at a red shift $z \leq 1.2$. The
 evolution of the equation of state $\omega_{\phi}$ and energy density $\rho_{\phi}$ with
 time is distinct from existing dynamical $\Lambda$ models based on slowly rolling fields,
 with $\omega_{\phi}$ and $\rho_{\phi}$ rapidly changing in a characteristic way
 following the transition. The $\phi$ energy density is composed of a time-dependent
 vacuum energy and coherently oscillating condensate component with a negative pressure.
 The condensate component will typically collapse to form non-topological soliton
 lumps, '$\phi$-axitons', which smoothly populate the Universe. 

\end{abstract}
\vfil
\footnoterule
{\small $^1$mcdonald@physics.gla.ac.uk}

\thispagestyle{empty}
\newpage
\setcounter{page}{1}

             Perhaps the most remarkable cosmological observation of recent times is the
  accelerating
 expansion of the Universe \cite{aex}. This requires the existence of an energy
 density which has a negative pressure and which is smooth on the 10Mpc scales relevant to
 dynamical estimates of the density of conventional dark matter. Other evidence for this
 comes from estimates of the age of the Universe, the Hubble constant, the baryon fraction in
 clusters, the galactic power spectrum and CMB measurements \cite{aex2,aex3}. The
 simplest explaination for a smooth energy density with a negative pressure is a time-independent
 cosmological constant, $\Lambda$. 
However, there are a number of problems with a fixed cosmological constant. 
The first is that it is difficult to understand why the cosmological constant should just be
 dominating the energy density at the present epoch \cite{pl1}. This has led to two different
 views. One is that the cosmological constant is due to the evolution of a scalar field
 ('tracking solution' \cite{trak}; see also \cite{wett}) whose energy density becomes significant recently due to
 dynamical effects. 
In general this approach has difficulties with nucleosynthesis and
 the present equation of state \cite{dod}. (A particularly promising version which may
 overcome these problems is given in \cite{muk}.) The other view is that
 the cosmological constant has become dominant in the present epoch due to anthropic
 selection (AS) \cite{pl1}. However, even if we assume that AS is responsible for the
 dominance of the 
cosmological constant, there are still problems. It is difficult to understand why a fixed
 $\Lambda$ is so small compared to the mass scales of particle physics ($\Lambda \sim
 10^{-120} {\rm M_{Pl}^{4}}$). This has led to the
 suggestion that the absolute minimum of the vacuum energy should be exactly zero and
 that the smooth energy density is due to the evolution of a scalar field towards
 the minimum, resulting in a {\it dynamical} cosmological constant, $\Lambda(t)$
 \cite{dyn1,dyn2,pngb,caldwell,plpot}. (An interesting alternatve based on quantum spinodial fluctuations has been suggested in \cite{holman}.) There is also observational evidence
 in support of a specifically dynamical cosmological constant \cite{obs1}. The amplitude of
 the COBE-normalized galaxy clustering power spectrum is too large in the case of a fixed
 $\Lambda$. However, 
if the effective cosmological constant is decreasing with time, then the amplitude of the
 power spectrum is reduced, alleviating the problem \cite{obs1}. 

           There have been a number of suggestions regarding the nature of the dynamical
 cosmological constant. Most popular are models based on pseudo-Nambu-Goldstone bosons
 (PNGBs) \cite{pngb,dyn1,dyn2,caldwell}, exponential potentials
 \cite{trak,dod,dyn1,caldwell} and
 inverse-power law potentials \cite{plpot,dyn1,dyn2}. All of these produce characteristic
 time-dependent energy densities
 $\rho_{\phi}$ and pressures $p_{\phi}\; (\equiv \omega_{\phi}\rho_{\phi})$ which come to
 dominate at recent red-shifts. For the PNGB models, the pressure slowly tends from
 negative values towards zero. For tracking models based on exponential and inverse-power
 law potentials, on the other hand, the pressure is evolving from a value typically positive and close to
 zero (corresponding to matter tracking) {\it towards} a negative value.
This should allow the models to be distinguished by
 precision CMB angular spectra measurements \cite{caldwell}. 
All of these models are based on very
 light scalar fields ($m_{\phi} \lae H_{o}$, where $H_{o}$ is the present value of the
 expansion rate) which are slowly rolling at the present time, $|\dot{\phi}/\phi| \lae H_{o}$. 

         In the present paper we wish to introduce an alternative model for a dynamical
 cosmological constant. This is based on the idea that the negative pressure energy density
 is associated with a conventional metastable false vacuum. This is perhaps the simplest form
 of negative pressure energy density in particle physics models. 
However, in order to have a dynamical cosmological constant
 we require that a phase transition from the metastable phase has
 occured recently. To achieve this we will consider the scalar field $\phi$ to have a time
 dependent mass squared term which has recently become negative. We will refer to this
 scheme as
 the 'very recent phase transition' (VRPT) scenario for a dynamical cosmological constant. 
The VRPT scenario is quite distinct from previous models based on slowly rolling
 scalar fields. In particular, there will be no need for the mass scale of the scalar to be
 extremely small, and the pressure and energy density will be rapidly evolving in a
 characteristic way at recent
 times following the phase transition. This should allow the VRPT scenario to be
 distinguished from the others by precision CMB measurements. 

             We will consider the usual spontaneous symmetry breaking potential for a 
real scalar field with $\phi \leftrightarrow -\phi$ symmetry\begin{footnote}{This can be
 generalized to a complex scalar. For a real scalar field 
domain walls may form during the VRPT, whilst for a complex field global strings may
 form.}\end{footnote}, 
\be{e1} V(\phi) = -\frac{\mu^{2}(t)}{2} \phi^{2} + \frac{\lambda}{4} \phi^{4} + \Lambda 
\;\;\;\;\;;\;\;\; \Lambda =  \frac{\mu_{o}^{4}}{4 \lambda}          ~,\ee
where 
\be{e2} \mu^{2}(t) = \mu_{o}^{2}\left(1 - \left(\frac{a_{c}}{a}\right)^{n}\right)   ~\ee
and $a_{c}$ is the scale factor at the time of the transition. In general the VRPT scenario
 requires an additional time-dependent $\phi^{2}$ term, which we will refer to as a
 'stabilizing interaction'. We will discuss some possible sources for the stabilizing interaction
 later, but for now we simply model it phenomenologically. We will see that such time-dependent mass squared terms with integer $n$ can arise naturally in plausible models. 

              Typically the mass scale of the potential will be very large compared with
 $H_{o}$. Thus following the phase transition $\phi$ will be coherently oscillating
 about
 the relatively slowly evolving time-dependent minimum of its potential. In order to
 discuss the time evolution of the
 $\phi$ energy density and equation of state, we follow the discussion of \cite{turner},
 based on averaging over the rapid oscillations of the scalar about the minimum of its
 potential.  

                The $\phi$ equation of motion is given by 
\be{e3} \ddot{\phi} + 3H \dot{\phi} = -\frac{\partial V}{\partial \phi}       ~.\ee
This may be rewritten as 
\be{e4} \frac{\partial}{\partial t} \left( \frac{\dot{\phi}^{2}}{2} +  V \right) = -3 H \dot{\phi}^{2} ~,\ee
where the partial derivative is with respect to constant $\mu^{2}(t)$.\begin{footnote}{In practice
 we evolve the energy density and pressure of $\phi$ by 
incrementing the energy density with $\mu(t)$ held constant and then incrementing $\mu(t)$ and 
calculating the pressure. Therefore we do not use the total derivative with respect to $t$
 given in \cite{turner} for an explicitly time-dependent potential.}\end{footnote}
Taking the time average over an oscillation cycle, we obtain
\be{e4} \frac{\partial \rho_{\phi}}{\partial t} = -3 H \gamma \rho_{\phi}   ~,\ee
where 
\be{e5}  \gamma = \frac{2 \int_{c} (1 - V/V_{max})^{1/2} d\phi}{\int_{c} (1 - V/V_{max})^{-1/2} d\phi}    ~.\ee
$\int_{c}$ denotes integration over one oscillation cycle and $V_{max} \equiv
 \rho_{\phi}$ is the maximum $\phi$ energy density during an oscillation cycle. 
$\gamma$ corresponds to the time average of $\dot{\phi}^{2}/\rho_{\phi}$ over an oscillation
 cycle. The equation of state of the $\phi$ energy density is then given by 
\be{e7} \omega_{\phi} \equiv p_{\phi}/\rho_{\phi} = \gamma - 1    ~.\ee
We have calculated the time evolution of the energy density
 and equation of state for different values of $n$ as a function of red-shift, $z$. The
 parameters of the potential are chosen such that we have at present $\Omega_{\phi}
 \approx 0.7$ and $\Omega_{m} \approx 0.3$ (where $\Omega_{m}$ is the density of conventional clustered dark matter). 

              The numerical solutions for $\rho_{\phi}$ and $\omega_{\phi}$ as a function of
 $z$ depend on $\mu_{o}^{2}$ and $\lambda$ only through the ratio
 $\mu_{o}^{4}/\lambda$; we have calculated the evolution for the case
 $\lambda = 1$. Because of this scaling property, for a given value of $\omega_{\phi}$ today, $\omega_{\phi\;o}$, the
 evolution of the $\phi$ equation of state and energy density is completely fixed by the
 value of $n$ in the stabilizing interaction.

         In Figure 1 we show the evolution of the equation of state $\omega_{\phi}$ as a 
function of red-shift for $n = 2,3,4$. In Table 1 we give the parameters of the VRPT
 ($\omega_{\phi\;o}$, $\mu_{o}$ (for $\lambda = 1$) and $z_{c}$, the red-shift at which
 the transition occurs) for the cases $n=2$ and $n=4$. 
\begin{figure}
\leavevmode
\centering
\vspace*{90mm} 
\includegraphics{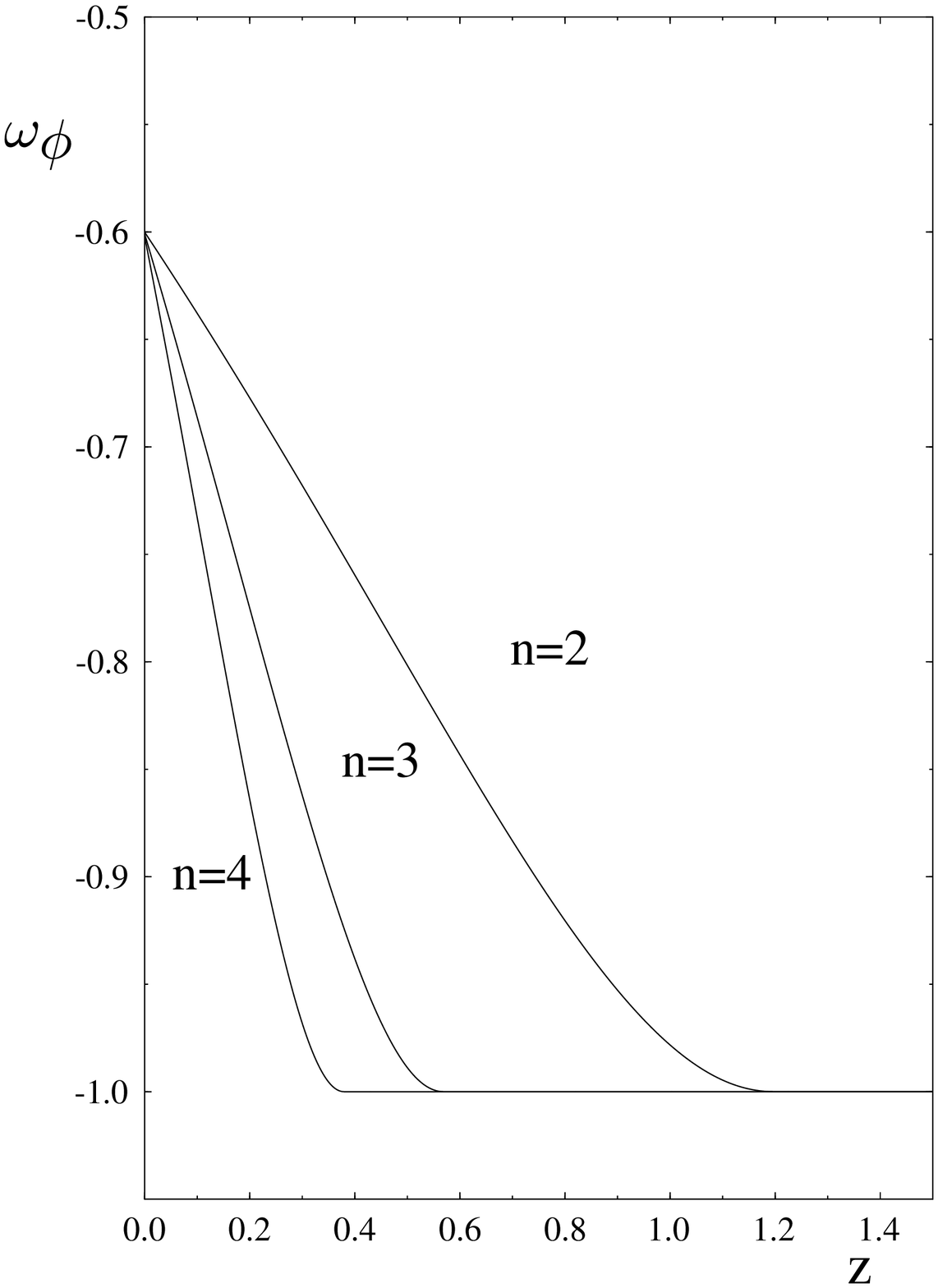}
\includegraphics{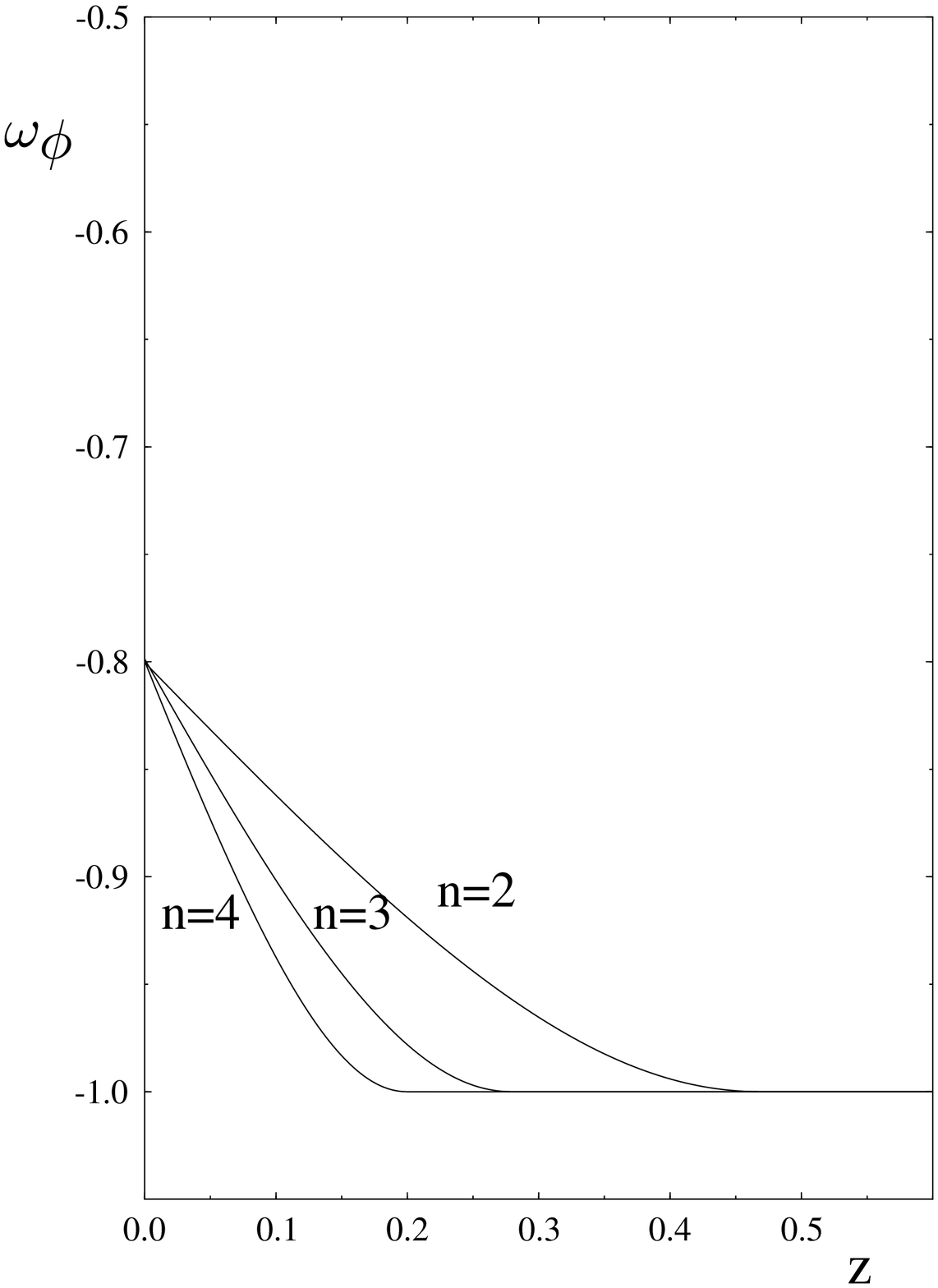}
\caption{Evolution of the equation of state for different $n$ with
 $\omega_{\phi} = -0.6$ and $\omega_{\phi} =-0.8$ at present.}
\label{kuva1}       
\end{figure} 
For $n \geq 2$ and $\omega_{\phi\;o} \leq -0.6$ we find that the phase 
transition occurs at $z_{c} \leq 1.2$. (We consider $\omega_{\phi\;o} \leq -0.6$, in keeping
 with observational limits for the case of a fixed $\omega_{\phi}$ \cite{wcos}. Since in our
 case $\omega_{\phi}$ is decreasing with $z$, this limit should be conservative.) 
From Table 1 we see that the mass of the scalar is $\lae 10^{-3} \eV$ for $\lambda \lae 1$.
In Figure 2 we show the evolution of the energy
 density together with the matter energy density $\Omega_{m}$ for the cases $n=2$ and
 $n=4$, where we have normalized the energy density by taking the ratio to the present
 critical density $\rho_{c}$. Both $\omega_{\phi}$ and $\rho_{\phi}$ rapidly change at
 recent red-shifts, the more so for larger values of $n$. 

         In addition, we have considered the effect of the VRPT on the age of the Universe, given by 
\be{7a} t_{U} = \frac{2}{3} H_{o}^{-1} f_{U} \;\;\;\; ; \;\; 
f_{U} = \frac{3}{2} \int \frac{da}{a} \left(\Omega_{\phi}(t) + \Omega_{m}(t) \right)^{-1/2}    ~.\ee
The age of globular clusters requires that $f_{U} = 1.5 \pm 0.3$ \cite{aex2,obs1}. For a
 fixed cosmological constant and $\Omega_{\Lambda} = 0.7$, $f_{U} = 1.45$. For the
 VRPT $f_{U}$ is generally smaller, but not significantly so. The largest deviation in the
 examples considered corresponds to $n=2$ and $\omega_{\phi\;o} = -0.6$, for which
 $f_{U} = 1.38$.
\begin{figure}
\leavevmode
\centering
\vspace*{90mm} 
\includegraphics{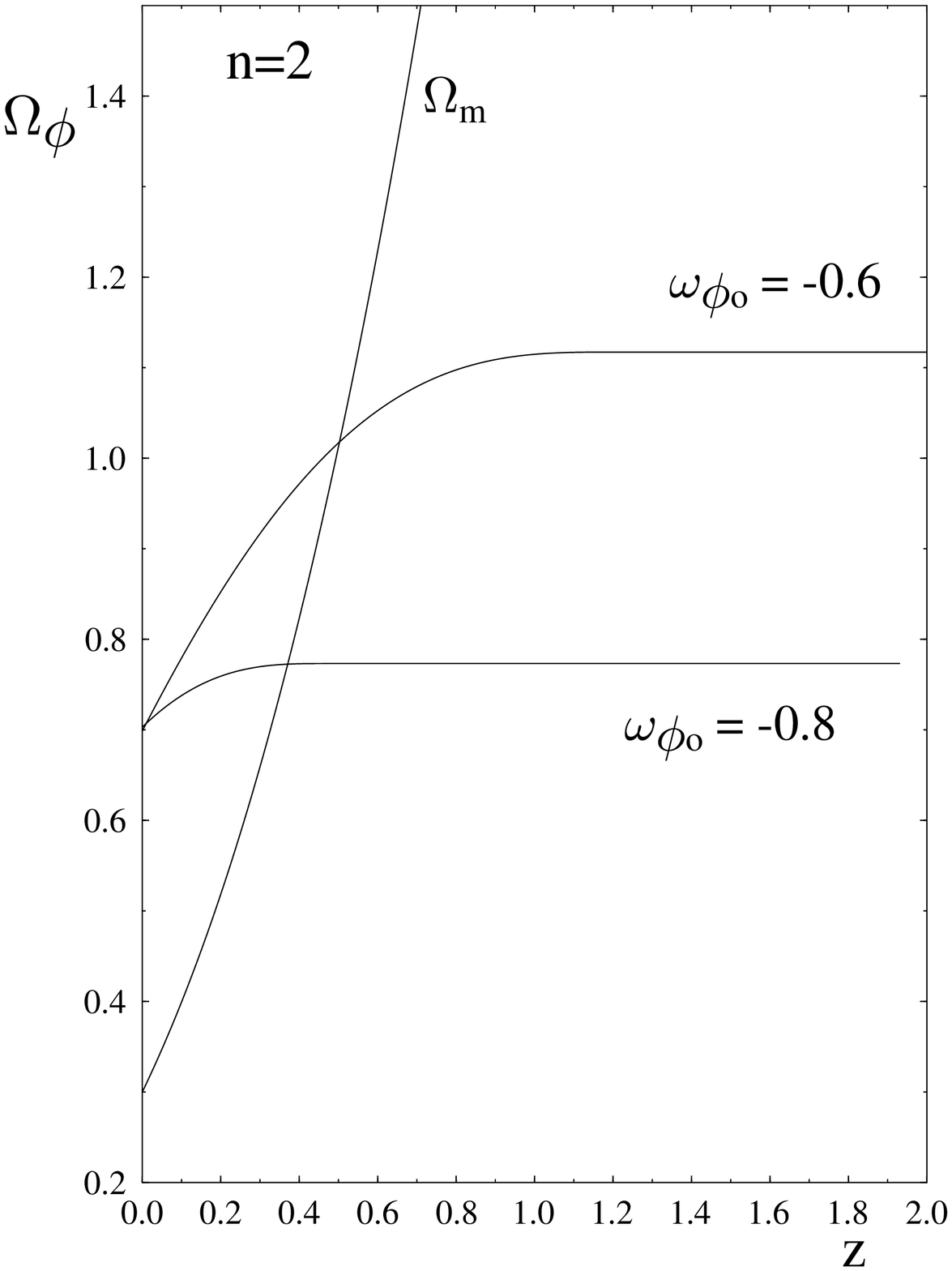}
\includegraphics{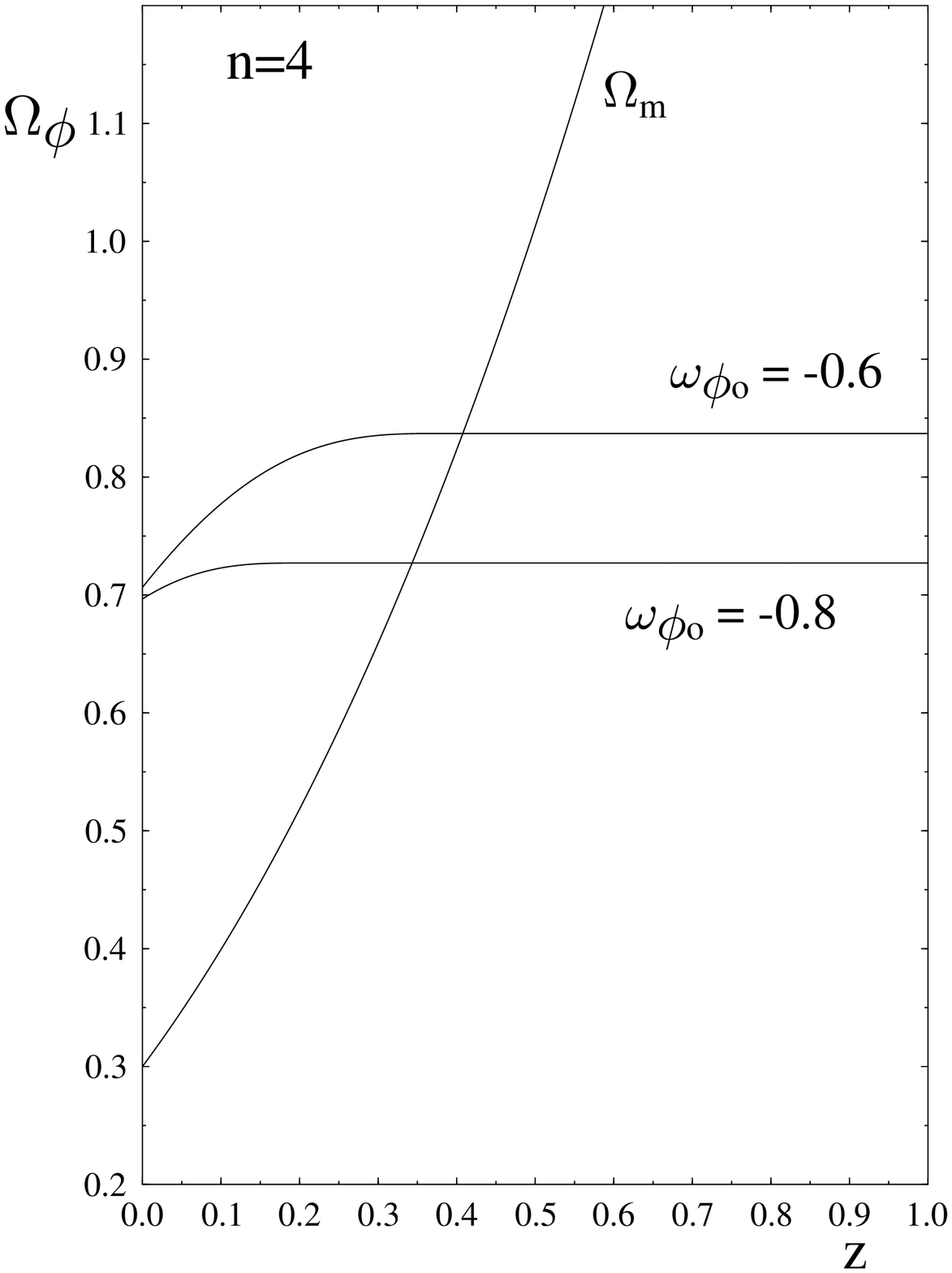}
\caption{Evolution of the energy density $\Omega_{\phi}(t) \equiv \rho_{\phi}(t)/\rho_{c}$ for $n=2$ and $n=4$.}
\label{kuva2}       
\end{figure} 
\begin{center} {\bf Table 1. VRPT parameters. } \end{center} 
\begin{center}
\begin{tabular}{|c|c|c|c|}          \hline
n & $\omega_{\phi\;o}$ & $\mu_{o}$ & $z_{c}$ \\ \hline
2 & $-0.6$ & $3.45 \times 10^{-3} \eV$ & $1.20$\\
& $-0.8$ & $3.15 \times 10^{-3} \eV$ & $0.47$\\
\hline
3 & $-0.6$ & $3.27 \times 10^{-3} \eV$ & $0.57$\\
& $-0.8$ & $3.13 \times 10^{-3} \eV$ & $0.28$\\
\hline
4 & $-0.6$ & $3.21 \times 10^{-3} \eV$ & $0.38$\\
& $-0.8$ & $3.10 \times 10^{-3} \eV$ & $0.20$\\
\hline
\end{tabular}
\end{center} 

               After the phase transition has occured\begin{footnote}{We are assuming that
 $\phi$ is out of thermal equilibrium, so that the transition is purely dynamical in
 nature.}\end{footnote}, the $\phi$ energy density will be composed of a time dependent
 vacuum energy $\rho_{vac}$ and a $\phi$ condensate component $\rho_{osc}$
 corresponding to coherent oscillations about the time dependent minimum of the potential.
 In Figure 3 we show the equation of state and energy density of the $\phi$ oscillations as a
 function of $z$. We see that the equation of state
rapidly tends towards the value $\omega_{osc} = 0$ corresponding to effectively
 $\phi^{2}$ 
oscillations about the minimum. Nevertheless, the equation of state and so the pressure is
 significantly negative throughout.  
In Figure 4 we show the equation of state and energy density associated with the
 time dependent vacuum energy, 
$\rho_{vac} \equiv V(\phi_{min}(t))$. We see that the equation of state does not
 significantly deviate from 
the value expected for a constant vacuum energy density, $\omega_{\Lambda} = -1$. 

\begin{figure}
\leavevmode
\centering
\vspace*{90mm} 
\includegraphics{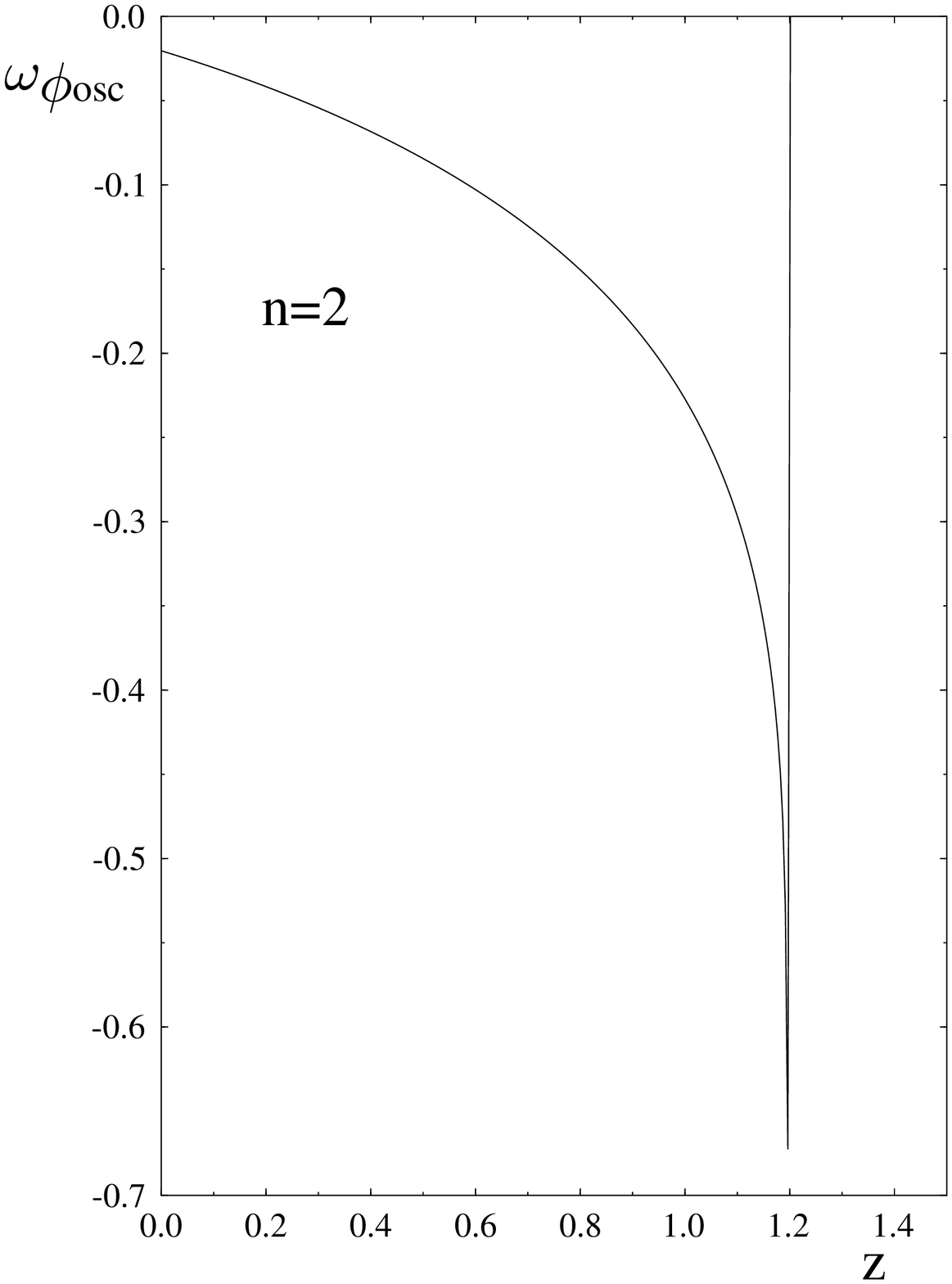}
\includegraphics{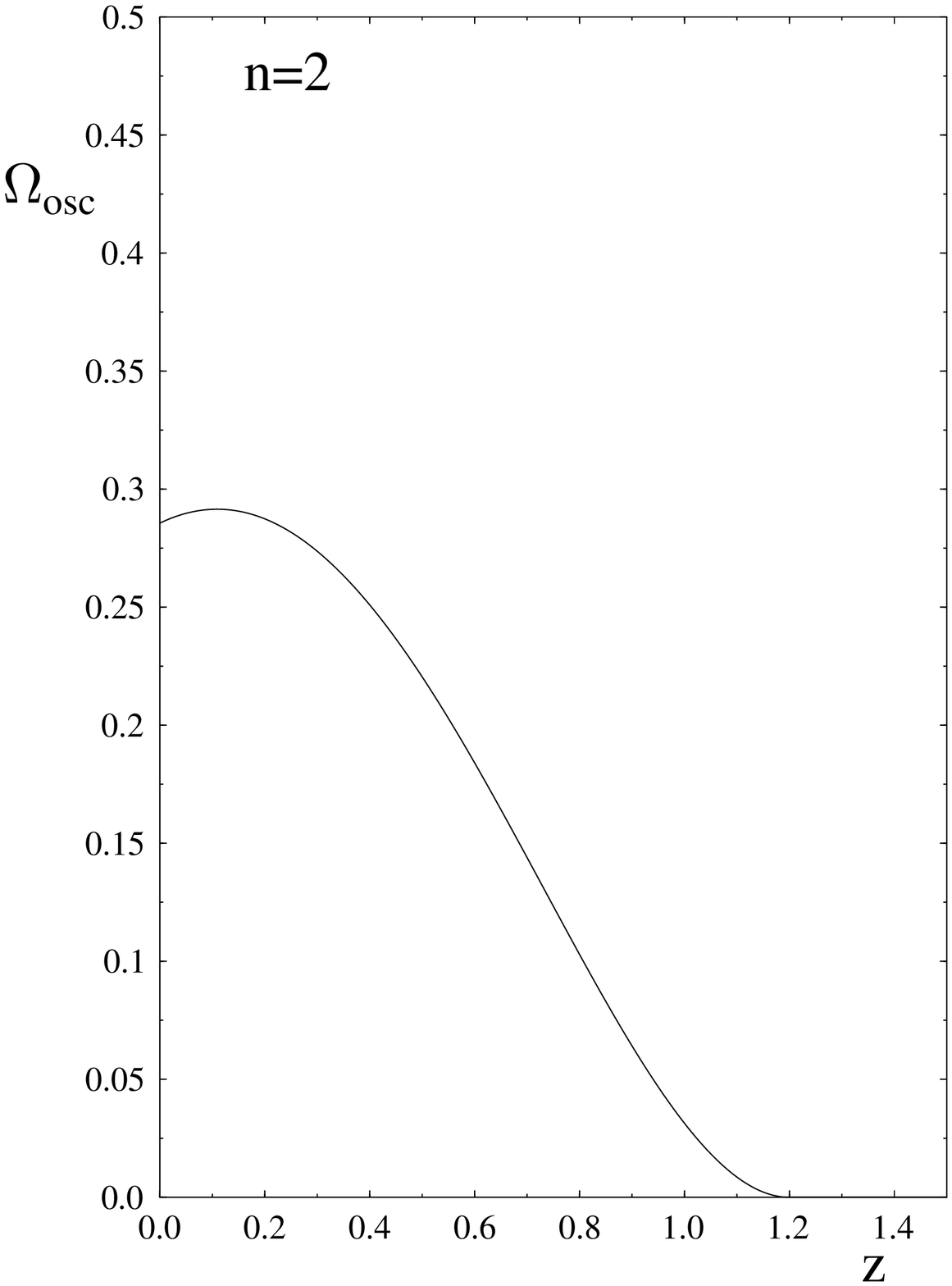}
\caption{Evolution of the equation of state and energy density of the $\phi$ coherent oscillations.}
\label{kuva3}       
\end{figure} 
\begin{figure}
\leavevmode
\centering
\vspace*{90mm} 
\includegraphics{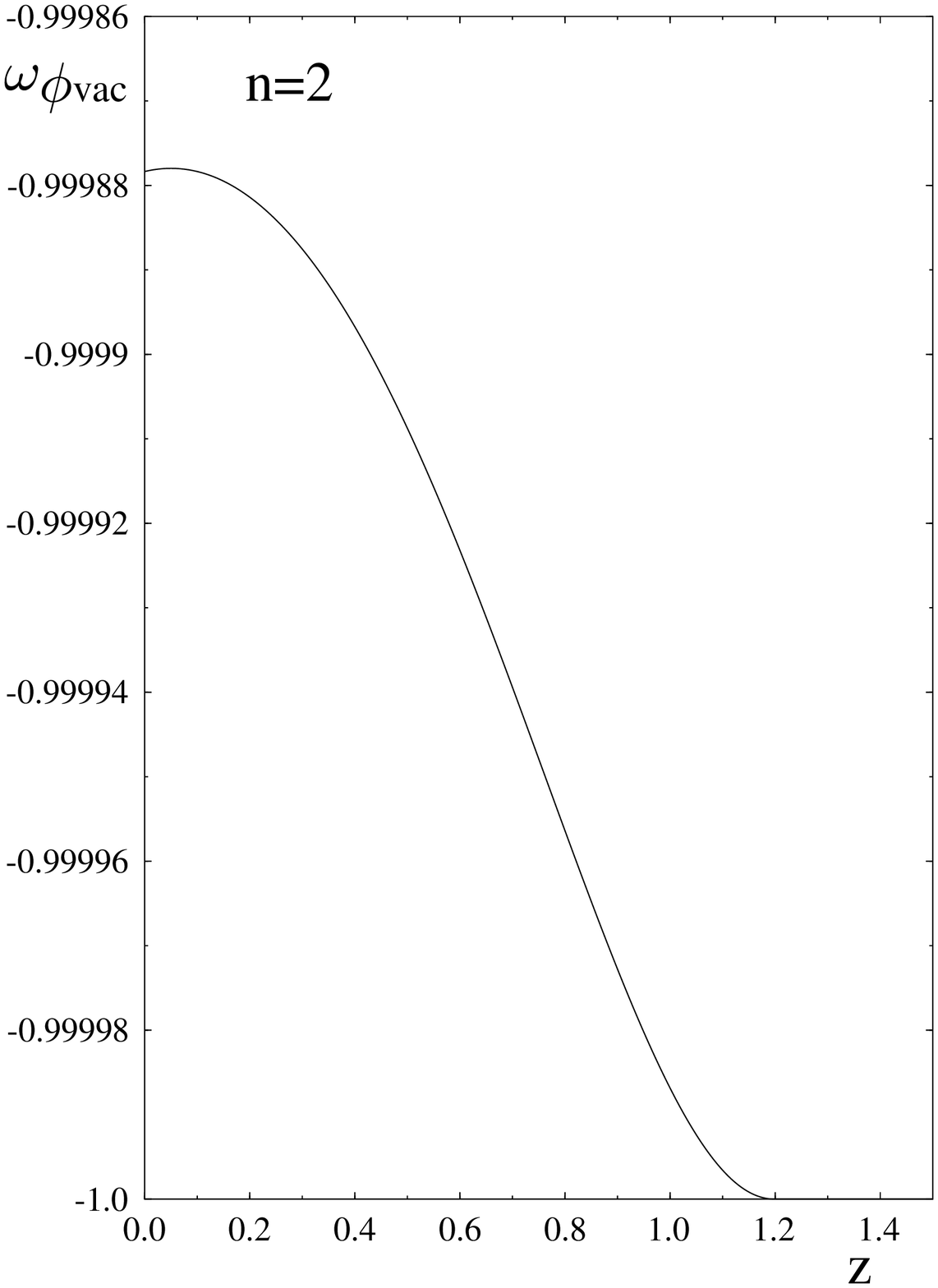}
\includegraphics{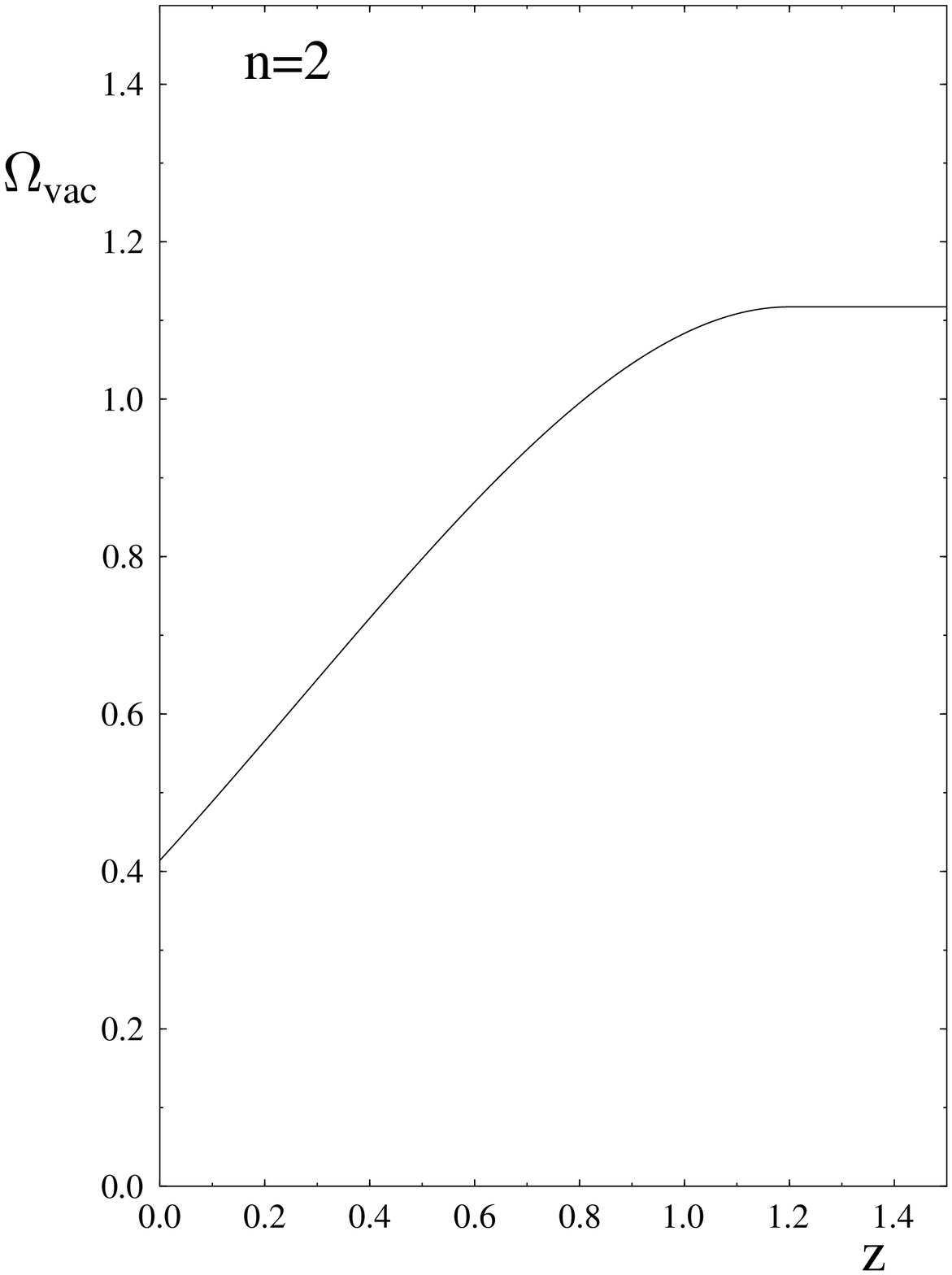}
\caption{Evolution of the equation of state and energy density of the time dependent vacuum energy.}
\label{kuva4}       
\end{figure} 

                  The negative pressure of the $\phi$ oscillations implies that the $\phi$ condensate
 is unstable with respect to spatial perturbations of $\phi$ \cite{ks1,ks2}. Such perturbations
 may be expected to exist, coming, for example, from thermal or inflationary quantum
 fluctuations. In the linear approximation the spatial perturbations evolve as
\be{e8}  \delta \ddot{\phi}_{\vec{k}} = |\omega_{\phi}| \vec{k}^{2} \delta \phi_{\vec{k}}  
~\ee
(where $\omega_{\phi} < 0$ is assumed). Therefore,
\be{e8a} \delta \phi_{\vec{k}} = {\rm exp}(|\omega_{\phi}|^{1/2} |\vec{k}| t) 
\delta \phi_{\vec{k}\;o}   ~.\ee
This is true so long as the wavenumber of the perturbation satisfies $|\vec{k}|^{2} 
\lae 4 |\omega_{\phi}| m_{\phi}^{2}$ \cite{ks2}, otherwise the positive pressure associated
 with gradient energy in the perturbation will overcome the negative pressure responsible for
 the growth of the perturbation. (For this reason, negative pressure effects play no role in 
the dynamics of $\Lambda(t)$ models based on slowly rolling fields, since their small mass,
$m_{\phi} < H_{o}$,  implies that perturbations on sub-horizon scales cannot grow.) The
 perturbations in the condensate will grow exponentially until they become non-linear. The
 condensate will then fragment into 
non-topological soliton lumps which we will refer to as '$\phi$-axitons' (following the
 existence of similar objects in axion cosmology \cite{axiton1} and Affleck-Dine
 baryogenesis \cite{axiton2}). The radius of the $\phi$-axitons will be determined by the
 first perturbation mode to go non-linear, $r_{\phi} \approx (|\omega_{\phi}|^{1/2}
 m_{\phi})^{-1}$ (assuming the exponential factor to be dominant in determining non-linearity). 
This will occur in a time $\delta t \approx (|\omega_{\phi}| m_{\phi})^{-1}{\rm log}(\phi/\delta \phi_{\vec{k}}) \ll H_{o}^{-1}$.

           Thus shortly after the VRPT the Universe will typically be filled with $\phi$-axitons. So long
 as $r_{\phi} \ll 10$Mpc, which is true if
 $m_{\phi} \gg 10^{-39} \GeV$, the $\phi$-axiton density will initially act as a smooth
 component of
 pressureless matter as far as 
determinations of the dark matter density are concerned. 

       We need to check that subsequent infall into galactic halos does not result in the $\phi$-axiton density clustering on 10Mpc scales and so no longer being effectively smooth. We do
 this via a simple Newtonian argument. We consider the mean distance between galaxies to
 be 
$R_{gal} \approx 10$Mpc, with $\phi$-axitons being smoothly distributed initially. The
 time scale for the $\phi$-axitons to fall a distance $\approx R_{gal}$ due to the attraction
 of a galaxy of mass $M_{gal}$ is then
\be{e9} t_{infall} \approx \frac{R_{gal}^{3/2}}{\sqrt{GM_{gal}}}     ~.\ee
Using $M_{gal} \approx 10^{11}M_{\odot}$ for the mean galactic mass, this gives
$t_{infall} \approx 1.6 \times 10^{12}$yr, which is much longer than the age of the
 Universe, $t_{U} \approx 10^{10}$yr. Thus infall will not significantly alter the
 smoothness of the $\phi$-axiton distribution. (The same argument holds for the $\phi$ particles in a homogeneous condensate.) Although we have estimated this at the present
 time, it holds for earlier times also, since $t_{U} \; (\propto H^{-1})$ and $t_{infall}
\; (\propto R_{gal}^{3/2})$ are both proportional to $a^{3/2}$.  

           Therefore in the presence of spatial perturbations, the energy being fed into the
 coherent 
$\phi$ oscillations will be converted into a smooth density of pressureless $\phi$-axitons.
 This will slightly alter the evolution of the total $\phi$ equation of state and energy density
 from the case where there are coherent $\phi$ oscillations with negative pressure. In 
Figure 5 we show the effect of replacing the $\phi$ condensate with pressureless 
$\phi$-axiton matter. The effect is a small alteration of the total $\phi$ equation of state and
 energy density as a function of $z$. 

\begin{figure}
\leavevmode
\centering
\vspace*{90mm} 
\includegraphics{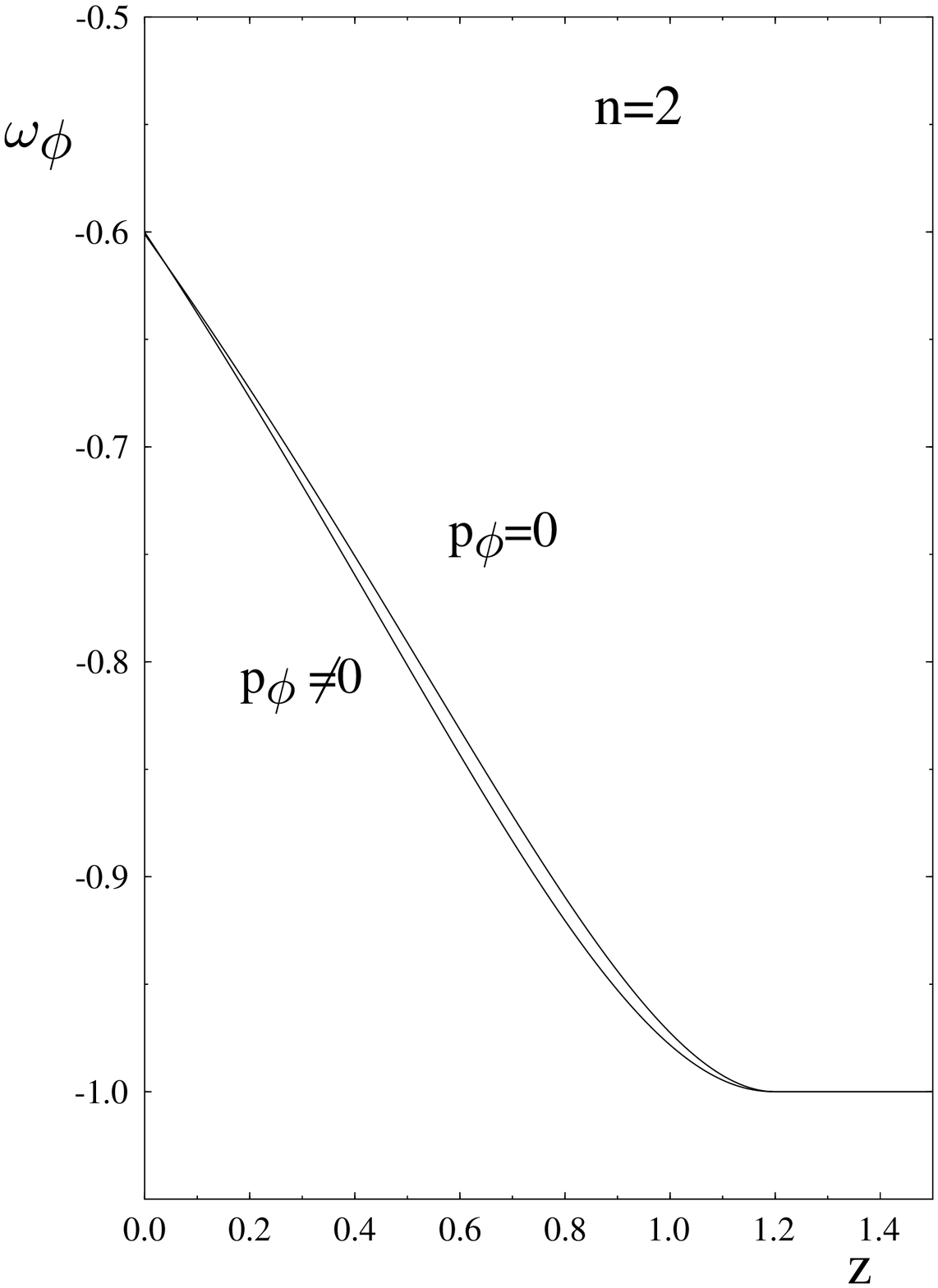}
\includegraphics{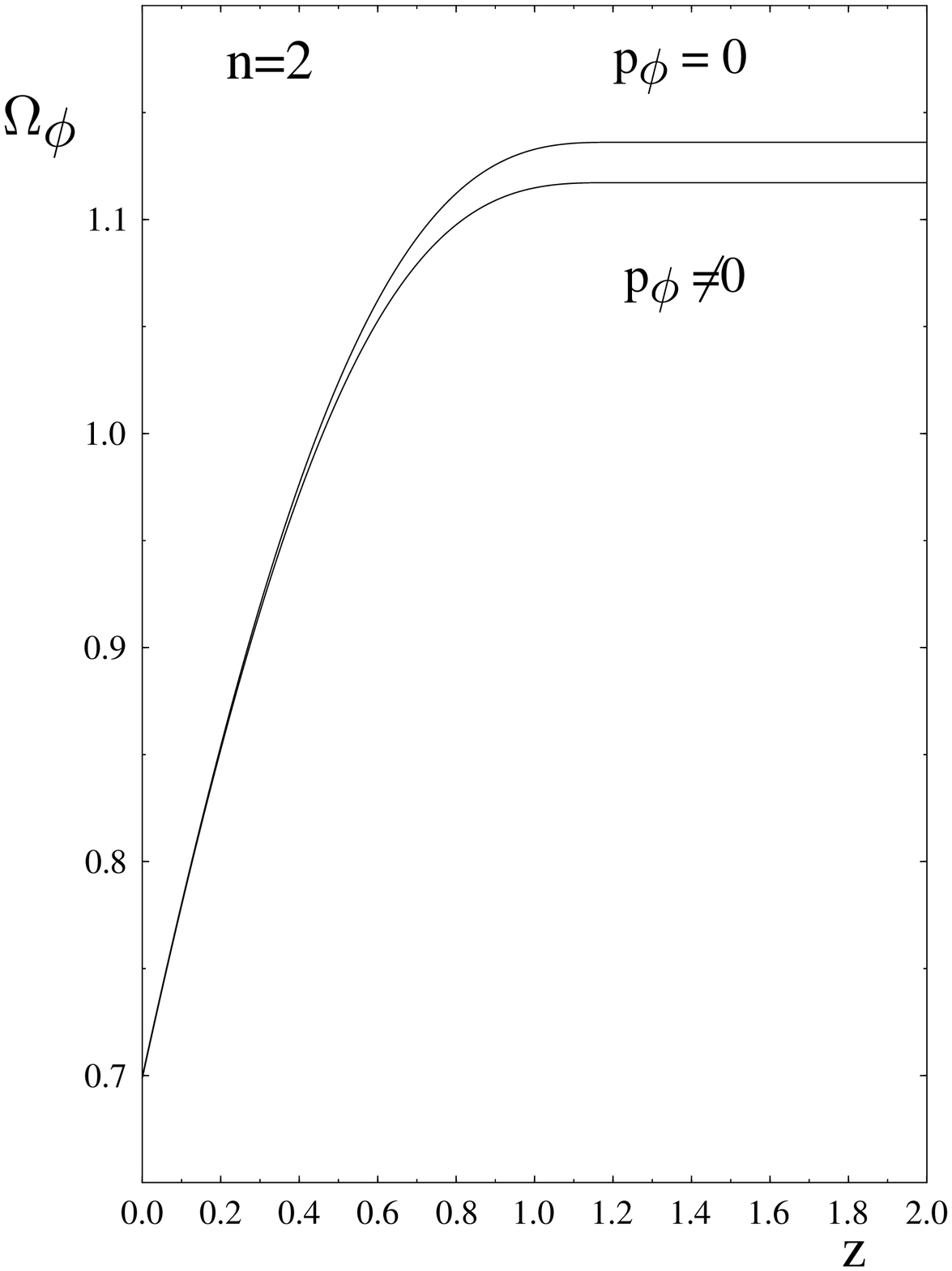}
\caption{Evolution of the total equation of state and energy density for the case of pressureless $\phi$-axitons.}
\label{kuva5}       
\end{figure} 
After the phase transition, the $\phi$ energy density consists of a time-dependent vacuum
 energy
\be{e10} \rho_{vac}  = \frac{\mu_{o}^{2} - \mu^{2}(t)}{4 \lambda}              ~,\ee
with $\omega_{vac} = -1$ to a very good approximation, 
and an energy density either in the form of a coherently oscillating $\phi$ field with a
 negative pressure or, more likely,  
in the form of pressureless $\phi$-axitons. However, the total $\omega_{\phi}$ is insensitive
 to the pressure in the condensate component, as seen from Figure 5. This means that
 $\omega_{\phi} 
> -1$ following the VRPT is simply due to the dilution of $\omega_{\phi}$ by the low
 pressure component of smooth $\phi$ condensate or $\phi$-axiton matter,
\be{e11} \omega_{\phi} = \frac{\omega_{osc}\rho_{osc} 
+ \omega_{vac}\rho_{vac}}{\rho_{osc} + \rho_{vac}} \approx           
\frac{-\rho_{vac}}{\rho_{osc} + \rho_{vac}}    ~.\ee

             So far we have introduced a simple time-dependent mass squared term ('stabilizing
 interaction') in order to trigger the phase transition and so produce a dynamical
 cosmological constant. We now consider some ways in which this term could be generated
 in particle physics models. One possibility is that the $\phi$ field might couple to a light
 field which has a thermal distribution, such as the photons or neutrinos. For example, one
 could consider a coupling of the form $\frac{\phi^{2}}{M^{2}} {\cal L}_{ke}$, where 
${\cal L}_{ke} = -\frac{1}{4}F^{\mu\nu}F_{\mu\nu}, \; \overline{\psi}\partial_{\mu}\gamma^{\mu} \psi$. On taking the average over thermal
 fluctuations of the fields in 
${\cal L}_{ke}$ we obtain an effective $\phi^{2}$ term $\sim \frac{T^{4}}{M^{2}}
\phi^{2}$. With $T \propto a^{-1}$ this results in a stabilizing interaction with $n=4$. 
Alternatively we might consider a very light additional scalar field $\chi$ with a thermal
 distribution (and a small enough energy density so as to avoid nucleosynthesis constraints),
 coupling to $\phi$ via a term $\chi^{2} \phi^{2}$. On averaging 
over thermal fluctuations of the $\chi$ field, this will give an effective $\phi^{2}$ term
 $\sim  T^{2} \phi^{2}$, corresponding to a stabilizing interaction with $n=2$. In both of
 these interactions we have effectively massless fields coupling to $\phi$, so that the time dependence of the 
stablilzing interaction is due to the red-shifting of the energy of the thermal fluctuations of
 the light fields. An alternative is to consider the case where $\chi$ is massive and is
 coherently oscillating in an effectively $\chi^{2}$ potential
about $\chi = 0$. We assume that the $\chi$-matter density is negligible compared with the
 conventional dark matter density. In this case the $\chi$ oscillation 
amplitude will be proportional to $a^{-3/2}$, resulting in a stabilizing interaction with
$n=3$. These are just a few examples; one could also consider, for example, $\chi$ to be a
 slowly rolling field which triggers the phase transition, analogous to what happens in hybrid
 inflation models \cite{hybrid}. 

                   There has recently been some discussion of the possibility of distinguishing between
 models with different time-dependent equations of state $\omega(z)$. In \cite{maor} it
 was suggested that in order to distinguish between models with $\omega(z)$ and
 $\omega = constant$ we would require an accuracy of less than 1$\%$ in determining the
 luminosity distance as a function of red-shift, $d_{L}(z)$, whereas 1$\%$ is regarded as an
 optimistic estimate of the accuracy of future experiements \cite{maor}. However, in \cite{weller}, it is suggested that with an appropriate fit to $\omega(z)$, $\omega(z) = 
\omega_{0} + \omega_{1}z + ...$, it is possible to distinguish between different models
 using the datasets expected from the SNAP satellite \cite{snap}. 
In particular, they show that a toy model with $\omega_{0} = -0.6$ and $\omega_{1} = 
-0.8$ can be clearly distinguished from constant $\omega$ models. Comparing with Figure
 1, we find that for $n=2,3,4$ the expansion parameters $(\omega_{o}, \omega_{1})$ are
 given by (-0.6,-0.33), 
(-0.6,-0.70) and (-0.6,-1.05) respectively. So at least for $n=3$ and $n=4$ the VRPT
 scenario should be clearly distinguishible by SNAP, and possibly for $n=2$ also, although
 this is not directly apparent from the results of \cite{weller} and requires further analysis.

          In conclusion, we have introduced an alternative model for a dynamical 
cosmological constant, based on the idea that a scalar field underwent a phase transition
 from a metastable phase at a very recent epoch, $z \leq 1.2$.  This VRPT scenario
results in a characteristic evolution of the equation of state and energy density which is quite
 distinct 
from the case of models based on slowly rolling fields, with the pressure rapidly rising from
 $\omega_{\phi} = -1$ and $\rho_{\phi}$ 
rapidly decreasing at very recent times. The solutions for $\omega_{\phi}(z)$ and
 $\rho_{\phi}(z)$ are uniquely determined by the present value of $\omega_{\phi}$ and 
the form of the time dependent mass squared term in the potential. 
Following the phase transition, the $\phi$ energy density will consist of a time dependent
 vacuum energy and a negative pressure $\phi$ condensate which typically fragments to a
 smooth pressureless density of non-topological solitons, '$\phi$-axitons'. 

                            As with most other
 dynamical $\Lambda$ models, it is implicitly assumed that the reason for the recent
 dominance of the $\phi$ energy density and the recent occurance of the phase transition is
 connected with anthropic selection. In general, dynamical $\Lambda$ models require two
 conditions; that the energy density in the scalar field has recently become dominant and that
 the scalar field energy density is varying significantly on time scales of the order of
 $H_{o}^{-1}$. Therefore two tunings are generally required. In this regard the VRPT
 scenario is on the same footing as other dynamical $\Lambda$ models. Ultimately the
 question of which dynamical $\Lambda$ model is correct is a
 matter to be decided by observations. A future goal will therefore be to understand
 the detailed predictions of this class
 of dynamical $\Lambda$ model for observable quantities; high-z supernova, quasar lensing
 statistics, the galaxy clustering power spectrum and in particular the angular CMB
 spectrum. We expect that these will be clearly distinguishable from other dynamical
 $\Lambda$ models, and indeed we have shown by comparison with recent analyses 
that at least some VRPT
 models are likely to be distinguishable by SNAP. In addition, the idea of a very recent phase
 transition 
leads to other interesting issues. One is whether the $\phi$-axiton density could be
 observationally or experimentally distinguished from conventional cold dark
 matter. This will depend on the $\phi$ mass and on how strongly the $\phi$ field interacts
 with ordinary matter. 
In addition, it is possible that the VRPT could result in the very recent formation of domain
 walls or 
global strings, depending on the symmetry associated with the potential and whether $\phi$
 is real or complex. It would be interesting to consider whether there are any observable
 effects associated with the very recent creation of topological defects. 
With the possibility of recently formed topological and non-topological solitons, the
 phenomenology of the VRPT scenario for the dynamical cosmological constant may be surprisingly rich.

\end{document}